# The development of a mapping tool for the evaluation of building systems for future climate scenarios on European scale

J.M. van der Steen & A.W.M. van Schijndel

March 4th 2015

# Contents

# 1. Introduction

A lot of studies exist of the performances of new innovative building systems. The performance of building systems is often dependent of external climate conditions. The variety of climate conditions in Europe make the performance of building systems dependent of the geographical location. Mapping the performance of systems on European scale can give more insight into the regions where building systems perform well or worse. Indicating in which region a building system performs well can be used for commercial purposes. Mapping building system performances on European scale decreases the chance that, due to choosing unfavourable climate conditions, innovations are unjustly classified to have a low performance.

The future climate conditions will differ from the current climate conditions [1]. Most building system are used for a longer period of time. Therefore is it important to consider the performance of building systems for a longer period of time. The purpose of this study is to develop a MATLAB model, where the performance of state-space based models can be visualised for Europe over an extended period of time. This MATLAB model will be usable for other users with knowledge of building systems and state-space models. The climate change will be evaluated using climate files created with the REMO model. Two state-space models of building systems will be used as an example to demonstrate the MATLAB model. These models are a heating, ventilation and air conditioning (HVAC) system and a concrete brickwork solar collector (SC) system.

## 2. Related work

This section will focus on important building related research areas which are relevant for the mapping of building system performances over an extended time period in the future. Literature has been reviewed concerning the impact of climate change on the performance of building systems. Literature has been obtained using Elsevier ScienceDirect. For the search of literature were combinations of the following keywords used: mapping, modelling, building systems, state space, model, climate change, Europe(an) and building performance.

The search for literature gave plenty results concerning energy demands of buildings and climate change. As stated in a review of the impact of climate change on energy use of buildings by D.H.W. Li et all [1], there is a growing concern about energy use of buildings and its implications for the environment. Energy usage and climate change influence each other. Due to global warming will the temperature rise and will there be a higher demand for cooling. More electricity usage for air conditioning would lead to larger emissions, which would increase climate change and global warming. D.H.W. Li et all [1] conclude that with a changing climate architects and building engineers can no longer assume a constant static condition for their designs, and need to consider the values of design variables for future years.

The literature review of D.H.W. Li et all [1] shows that a lot of researchers focus on the heating demand, cooling demand and peak loads. Their main findings were that in central and north Europe the decrease in heating, due to climate warming would dominate. In southern Europe would the climate warming and the increase in cooling and electricity demand outweigh the decreasing need for space heating. A literature review done by de Wilde and Coley [2] has a broader assessment on the implications of a changing climate for buildings. They state that the influence of climate change on buildings can range from gradual changes (for example the rise of average ambient temperature) to more extreme events (for example flooding). De Wilde and Coley [2] conclude that on building system level less research has been conducted on the impact of climate change. They mention that due to the life expectancy of building systems and the ongoing development of more efficient systems, research to performance of building systems for future climate change is difficult. Present day building systems are unlikely to still be used in the year 2050 without being updated or even replaced. Nevertheless do de Wilde and Coley [2] state that there is a need to conduct more climate change impact studies, to cover a wider range of systems and further climate scenarios for additional locations.

When searching for literature about the impact of climate change on building systems, the results are more scarce as established by previous mentioned literatures [1, 2]. For the assessment of the influence of climate change on buildings and building systems is a prediction necessary of the future climate conditions. Most building related studies are performed by using transient building simulation tools that require an hourly dataset to represent a year of climate data. Researchers (P. Xu et all, 2012) [3] that applied hourly climate data sets stress caution in the interpretation at such high temporal resolution. The methods show good agreement climatically (for example 10-year mean values), but hourly results are viewed with concern. Therefore can hourly climate datasets be best used for assessment of gradual changes and not for the extreme events.

J. Wachsmuth et all [4] have assessed the impact of climate change on solar and wind power generation for Germany's Northwest Metropolitan Region. For their analyses the created climatic values by using two regional climate models available for Germany. The first model is REMO, developed at the Max-Planck-Institut fuer Metereologie. The second climate model is the CLM model developed by the consortium of BTU Cottbus, Forschungszentrum GKSS, and Ptsdam-Institut fuer Klimafolgenforschung. To incorporate uncertainties in future climate development they created two time periods of 30 years. They created mean values of the two climate models for the 30 year time periods to observe the evolutions of selected climate parameters. The climate parameters of the different time periods are applied for models which asses the performance of solar power generation and wind power generation. This results in the gradual change of performances of these renewable power sources due to climate change, which can be compared with historic data.

A more detailed and full assessment of the near future change in productivity change of photovoltaic energy (PVE) using future climate variables has been performed by M. Gaetani et al [5]. The climate variables for this research are created by the ECHAM5-HAM aerosol-climate model. With a model for the performances of photovoltaic systems is an assessment done of the performance for the multiple time periods. The performance has been evaluated by comparing the mean performance of the year 2000 (present) and the year 2030 (future). This has been done for the climatic parameters to compare the differences between mean annual values. The performance has been determined as the percentage difference of the energy performance of the PVE. The performance has been determined for Africa and Europe in one image. Mapping the performance values in images as colours is used to evaluate the performance.

Ordinary differential equations (ODE's) are traditionally applied in computer simulations to evaluate (building)systems. As demonstrated by F.T. van Schie [7] and A.W.M. van Schijndel and H.L. Schellen [8] can (building)systems also be simulated with state-space models. The state-space functions are preferred, due to a reduced calculation time needed to solve them. For his research has F.T. van Schie [7] described a wind turbine and a concrete brickwork solar collector with state-space models to evaluate their performance in the Netherlands. With the usage of hourly data from the weather stations retrieved from the online KNMI (Royal Netherlands Meteorological Institute) database and by defining a performance indicator is per location a single value determined. With this method is the performance of a system not evaluated on single hourly results, but rather on averages or statistics of the results.

The research of A.W.M. van Schijndel and H.L. Schellen [8] has applied a similar approach on a larger geographical scale. Their research focused on a European scale instead of only on the Netherlands. For the mapping on European scale are over 130 external hourly-based climate files used, which were produced using commercially available software (Meteonorm 2011). The current hourly-based model HAMBase, part of the Heat, Air and Moisture Laboratory (HAMLab, 2013) has been used for the whole building model. To evaluate the performance of the building performance simulation tool, the approach Bestest (ASHRAE, 2001) has been used. The evaluation is performed by comparing results of the tested tool with results by reference tools. The MATLAB model has been used to show results for three cases: the indoor climate, heating and cooling. The main performances of a reference building model subjected to EU weather stations date, were visualized using EU maps. By changing a single building component, the single effect of this change is visualized by subtracting the new results with the reference case results. The evaluation of this method is also based on statistics and not on single hourly results.

## 3. Goal and outline

In the previous section are several possibilities shown of the application of mapping. Research of the performance of building systems is scarce. This is due to the fact that building systems have a relatively short lifetime (20-30 years) compared to the time needed for significant climate change to take place. M. Gaetani et al [5] have done an assessment to the performance of photovoltaic energy. Their research is limited for this specific system and results can't be applied to other systems. The research paper by F.T. van Schie [7] has focused mainly on an evaluation tool for different building systems in the Netherlands. The paper by A.M.W. van Schijndel and H.L. Schellen [8] has mainly focused on an evaluation tool for the performance of the indoor climate performance of buildings on European scale. Both researches focus on an evaluation tool for buildings or building systems, but neither focus on the impact of future climate on buildings or building systems.

The goal of this project is to develop a tool for the mapping of the performance of building systems on European scale for different (future) time periods. The tool should be easy to use for users and be applicable for different building systems. Users should also be able to use a broad range of climate parameters to assess the influence of climate change on these climatic parameters. Also should the calculation time be reasonable short. The mapping tool is developed in MATLAB, which can be used by other users for their own studies.

The model built-up is discussed in section 4. This section will discuss the overall developed MATLAB-files of the model, the usage of state-space models and the used climate files. Also is discussed which aspects of a building system should be defined in the model.

In section 5 is a discussion of the developed tool and if it meets the desired requirements defined as a goal for the tool. Conclusions are drawn in section 6.

In section 7 are three cases shown: climate change, a HVAC system and a solar collector. These cases are used as an example to demonstrate the usage of the tool for different cases and the obtained results. The first case shows how climatic variables change due to climate change. For the other two cases shall the state-space model be shown and the performance indicators defined for two building systems.

# 4. Model built-up

## 4.1. General set-up of MATLAB files

This paragraph will give general information of the built-up of the MATLAB model. The MATLAB model has been created with the MATLAB version: MATLAB R2014a. For the MATLAB model are the following toolboxes used: Control System Toolbox and Mapping Toolbox. In figure 1 is an overview given of the created MATLAB files and MATLAB functions to calculate the performance of an user defined building system dependant of climatic data from different time periods.

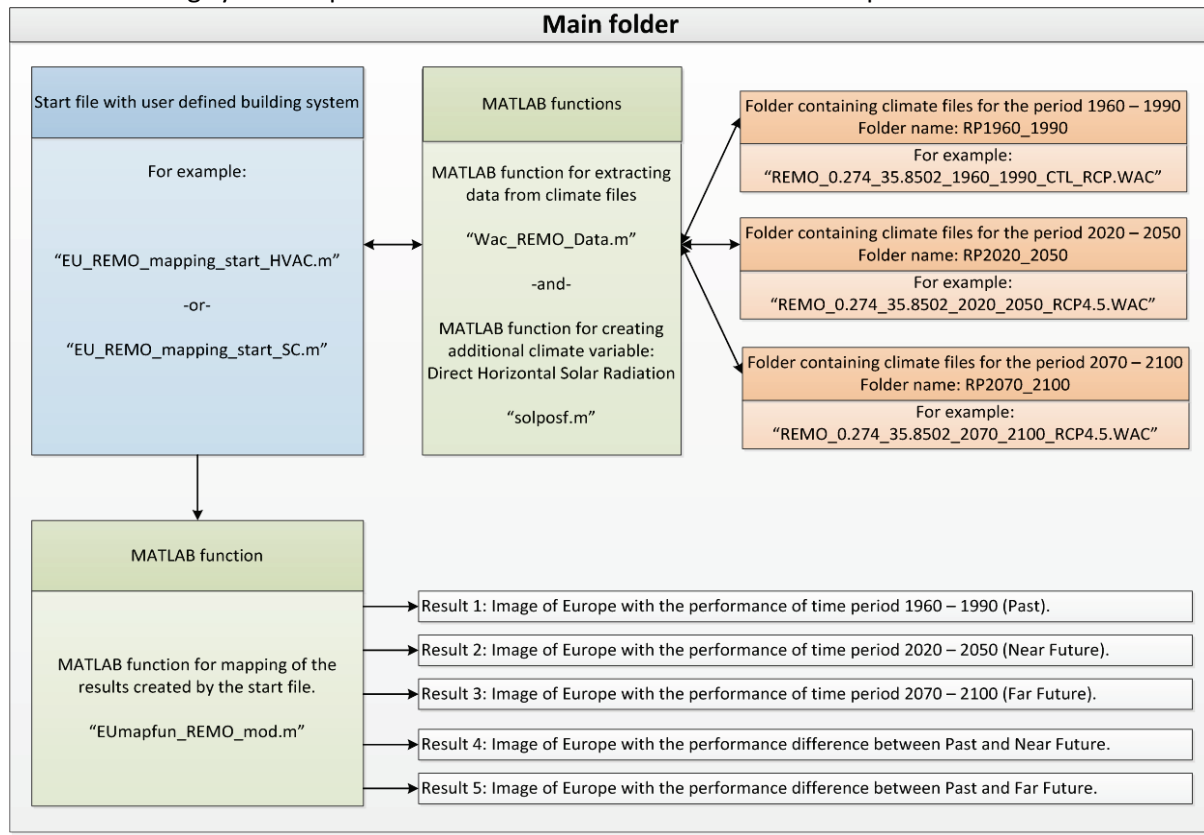

**Figure 1: Overview of created MATLAB files, MATLAB functions, used climatic data and results.**

In the blue box is the start file shown. This is the only MATLAB file which should be altered by users for their specific building system. In the start file are the following aspects of the systems defined: constants, state-space model, start values, used external climate data and the performance indicator. When the start file is filled in correctly, it can be run to create the results.

In the green boxes are MATLAB functions shown. These functions are used by the start file for two purposes. Two functions are used to create climatic variable. One function is used to extract climatic data from the climate files. The climate files are shown in the red boxes. Per time period is a folder used which contains the climate files of all the locations for that time period. Another function is used to create an additional climate variable: direct horizontal solar radiation [8]. A third MATLAB function is used to convert the created results of the start file in figures of European maps. In total are five maps created, where the performance values are shown as colours in an image of Europe. Three images are created to present the performance of the three time periods in maps. Two maps are created to show the difference of the performance of the two future time periods and the past time period.

## 4.2. MATLAB start file

The MATLAB start file contains a model of the building system of which the performance is mapped. The building system model can be named, which will be used to generate unique filenames to save results. In the MATLAB model are the following aspects of the systems defined: constants, state-space model, start values, used external climate data and the performance indicator. This information is used to determine a performance value for the three time periods for each location. The start file contains a loop to calculate performance indicators for the three time periods. Within this loop is another loop defined to calculate the performance indicators of each geographical location. A schematic overview of the file is given in figure 2.

### 4.2.1. State-space model

As described by F.T. van Schie [7] can the calculation time be reduced by using state-space models instead of ordinary differential equations. State-space is derived from the state-variable method, which describes a dynamic system with a set of first-order differential equations in the vector-valued state. State-space combines the set of first-order differential equations into matrices, which provides the possibility to work directly with the state-variable description of the system. In equation 1 are the formulas for the state-variable form shown. In the MATLAB model are the system matrix (A), the input matrix (B), the output matrix (C) and the direct transmission term matrix (D) defined.

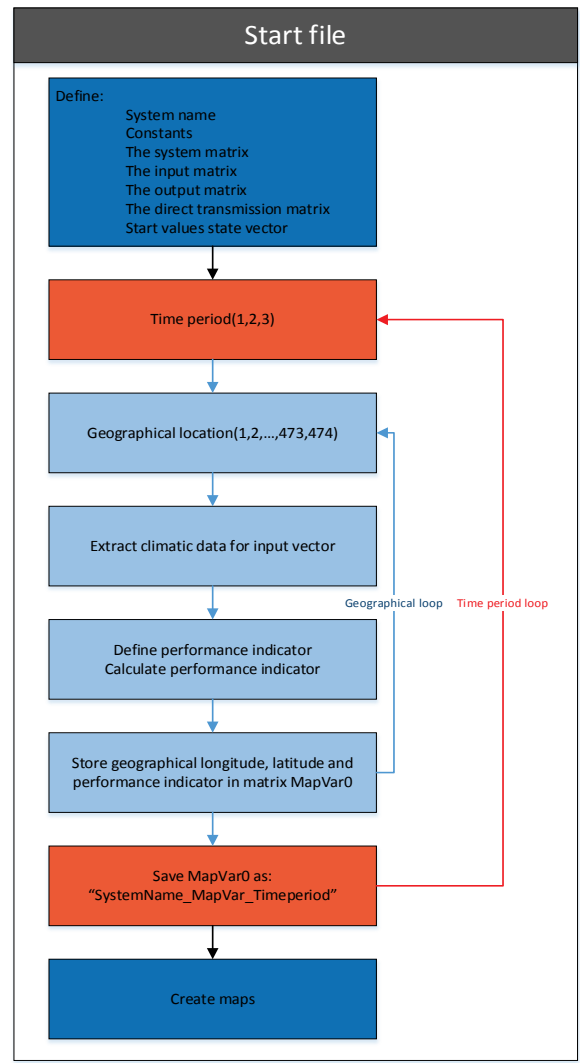

**Figure 2: Schematic overview of the start file built-up**

**Equation 1: The equations of the state-variable form represented as the vector equation, from van Schie, FT (2013).**

$$\dot{x}(t) = A \cdot x(t) + B \cdot u(t)$$
$$y(t) = C \cdot x(t) + D \cdot u(t)$$

With:
x = the state vector
y = the output vector
u = the input vector

A = the system matrix
B = the input matrix
C = the output matrix
D = the direct transmission matrix

For the state-space model do the following aspects need to be defined: constants, state-space matrix and the start values of the state vector. The system matrix (A) and the input matrix (B) are used to define the building system. Both matrices only contain known constants, therefore are the constants defined first in the MATLAB start file. The system matrix (A) and the input matrix (B) are designed to define the system with the state vector and the input vector.

The output matrix (C) and the direct transmission matrix (D) are designed to create an output vector. When for instance the aim of the research is to determine the state vector, than should the output matrix be an identity matrix and the direct transmission matrix should only contain zeros. These matrices can be used to create the state-space model, by using the command "ss(A,B,C,D)".

The state vector is time dependent; a new state vector value is calculated from the previous state vector value. Therefore are start values necessary for the state vector, the initial conditions. The input vector describes the climate variables and is also time dependent. The input vector is created with climatic data created using the REMO model [9] . These need to be defined in the MATLAB start file, to make it possible to create an output vector.

### 4.2.2. Input vector created with climatic data

The input vector u(t) contains climatic variables. The climatic data is extracted from climate files. The EU FP7 project 'Climate for Culture' [9] has made the climate files available, which are used for this model. This project develops climate files for three periods: 'the past' (1960-1990), 'the near future' (2020-2050) and 'the far future' (2070-2100) using the REMO model (Jacob et al. 1997) and a moderate climate scenario. For each time period there are climate files available for 474 locations in Europe, as shown in figure 3. The density of locations is higher than shown in the previous researches [7, 8] in section 2. Each climate file contains geographical information and external climate data.

The following geographical information is given in the climate files: Longitude [°], Latitude [°], Height above sea level [m], Time Zone [h from UTC] and Time Step [h].

The following external climate data are available to use for the model: Air temperature [°C] (TA), Relative Humidity [%] (HREL), Horizontal Global Solar Radiation [W/m²] (ISGH), Diffuse Horizontal Solar Radiation [W/m²] (ISD), Air pressure [Pa] (PSTA), Rain intensity [mm/h] (RN), Wind Direction [°] (WD), Wind Speed [m/s] (WS), Cloud Index [-] (CI), Atmospheric Counter Horizontal Long wave Radiation [W/m²] (ILAH), Atmospheric Horizontal Long wave Radiation [W/m²] (ILTH), Ground Temperature [°C] (GT), Ground Reflectance [-] (GR) and Direct Horizontal Solar Radiation [W/m²] (ISvar).

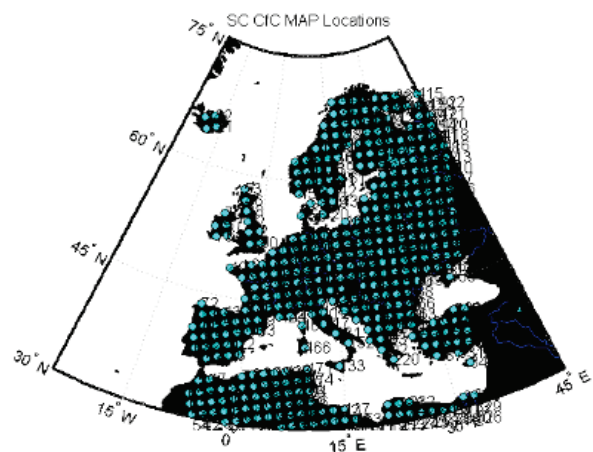

**Figure 3: Locations of the simulated climate files using the REMO model.**

For three time periods will the performance be mapped: 'the past' (1960-1990), 'the near future' (2020-2050) and 'the far future' (2070-2100). The input vector and therefore the performance indicator are both dependent from the used climatic files. The input vector and the performance indicator are defined in a loop. The loop will repeat the same calculations for the three different time periods using the therefore corresponding climate data. Within this loop is another loop created, to calculate the output vector and the performance indicator for each geographical location (figure 2).

The climate data is automatically extracted from the climate files with the MATLAB function file "Wac_REMO_Data.m". Within this MATLAB function is another MATLAB function used, created by A.M.W. van Schijndel [8], to create an additional climate variable: direct horizontal solar radiation. The MATLAB function "Wac_REMO_DATA.m"  imports first the climatic data file and separates the header and data. Geographical values are subtracted from the header text using the "sscanf" command and temporary stored in a 1-by-6 vector. The climatic data values are temporary stored in a 271560-by-14 matrix.

### 4.2.3.  The performance indicator

The performance indicator is already partly defined in de state-space model by the output matrix (C) and the direct transmission matrix (D), creating the output vector. An output vector is created for each time step in the climatic data. The output vector is calculated with the command "lsim", which simulates the time response of a dynamic system to arbitrary inputs. The state-space model, the input vector, the system time units and the initial conditions are used as input for the command "lsim". This will result in a 271560-by-1 output vector, which is time-dependent. Only a single value per location can be mapped in an image. Therefore is it necessary to transform the output vector in a single performance value. This can be done by creating mean values or statistic values. This will also reduce the risks of the low accuracy of hourly values [3]. Results are stored in a temporary matrix named "MapVar0" (Mapping variables). For each location is the longitude, the latitude and the single performance value placed in the matrix "MapVar0". For each time-period is a "MapVar0" matrix created, which is saved under unique names, containing the defined system name and the time period.

The performance indicators are created for each time-period, these are combined in one matrix, named: "MapVar". This matrix contains of each geographical location the following data: longitude, latitude and five different performance values. The first three performance values are the single performance values for each time period: the past, the near future and the far future. The fourth performance value is the difference between the near future and the past. The fifth performance value is the difference between the far future and the past. Due to the possibility of a small difference in performance between different time periods, two extra performance values are created. Similar as done by A.M.W. van Schijndel and H.L. Schellen [8], will the results of the past time period be subtracted from the two future time periods. One value represent the difference in performance between the time periods 'the past' and 'the near future'. The other value represents the difference in performance between the time periods 'the past' and 'the far future'.

## 4.3.  Mapping of results

The matrix "MapVar" is used by the MATLAB function "EUmapfun_REMO_mod.m" as input to create five maps. The figure names are defined as: "Past", "Near Future", "Far Future", "Difference Near Future and Past" and "Difference Far Future and Past". The MATLAB function "EUmapfun_REMO_mod.m" is a modified version of the MATLAB function "EUmapfun_mod.m" created by A.M.W. van Schijndel [8]. Only small modifications are made to create five figures with the proper figure names.

For each geographical location is the performance shown as a colour. Between geographical locations is the performance interpolated, thus creating fully coloured image of Europe. The differences of the climate of different time periods is small, therefore is the performance of systems often very similar with in percentage a small deviation. This creates very similar images of the performance of the different time periods, as can be seen in images 4,5 and 6, which show the example of a concrete brickwork solar collector. The example of the concrete brickwork solar

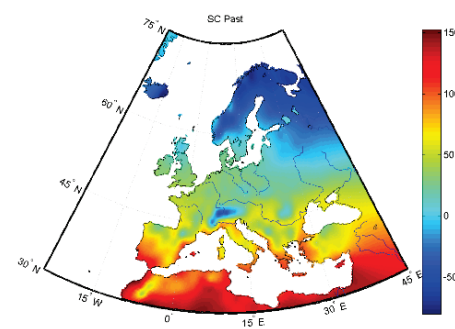

**Figure 4: The mean performance of the solar collector during the period 1960-1990.**

collector will be fully discussed in section 7.3.

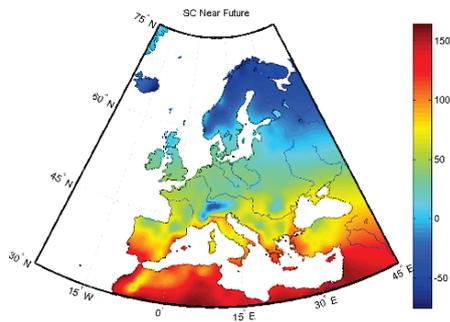

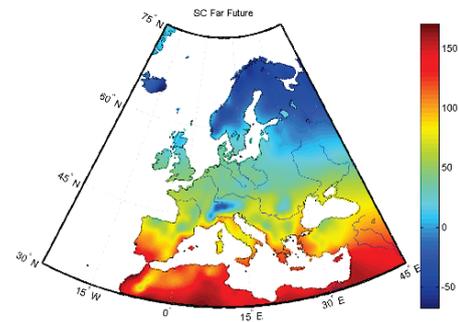

**Figure 5: The mean performance of the solar collector during the period 2020-2050.**

**Figure 6: The mean performance of the solar collector during the period 2070-2100.**

The images of the "Past" subtracted of the "Near Future" and the "Far Future" gives a more clear representation of the effect of climate change on systems. This is shown in images 7 and 8, which show the example of a concrete brickwork solar collector. As can be seen in the images are the colour differences more distinguishable, due to the smaller range of values that is mapped. These images are more suitable for the assessment of the effect of climate change on building systems.

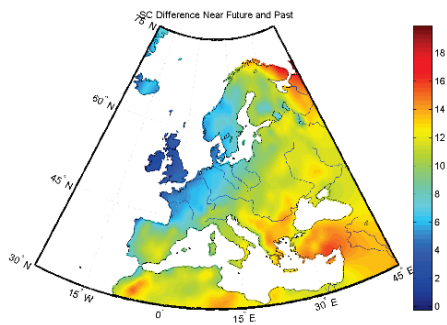

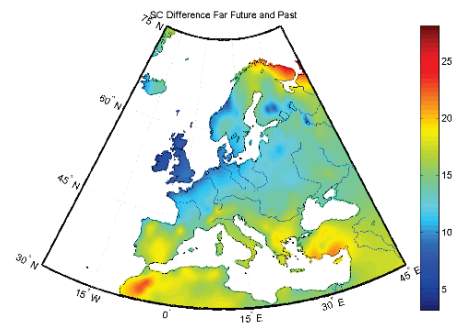

**Figure 7: The difference in performance of the solar collector between the periods [2020-2050] – [1960-1990].**

**Figure 8: The difference in performance of the solar collector between the periods [2070-2100] – [1960-1990].**

## 5. Discussion

The goal of this project was to create a model which can easily be used to determine the effect of climate change on the performance of building systems. A basic representation of a building system as a state-space model can suffice for the evaluation of the performance. The performance evaluation is limited to mean values or statistics, due to the limitation of the mapping of single values. Often is the accuracy of simulated single hourly values viewed with concern [3], but is the simulated gradual change of climate data over an extended time period often viewed as more accurate. Therefore will the assessment of mean values or statistics of a time period of 30 years lead to more trustworthy results.

When the performance of a building system is being mapped for different time periods the differences in performance are often small, thus creating similar images per time period. The results of a time period ('the past') should be subtracted from other time periods to make the difference more visible. This will directly show the effect of climate change directly in a more clear image.

For the results is interpolation used for areas between locations with climate files. Due to the high density of locations with climate files, is this still an accurate method. In Figure 3 is shown that an area east/south from Europe has no climate files to determine the performance. Therefore are the results for this area more untrustworthy and possibly untrue.

Due to the high amount of climate files (474 locations for 3 time periods) does the model need time to calculate the performance of each location for each time period. The time needed for the model to calculate all the values is dependent of the computational power. The shortest calculation time took approximately one hour.

## 6. Conclusion

It is possible to evaluate the change in performance of building systems over an extended time with a MATLAB model. By using three different time periods, the differences in performance can be compared. By plotting the difference in performance directly in an image, the increase or decrease in performance is more visible.

## 7. Examples

The model has been used for three cases. These cases are presented in this section as an example to demonstrate the usage of the model. The three cases are an evaluation of the climate change, a HVAC system and a brickwork solar collector.

### 7.1.    Climate change

A subject of this research is the climate change in Europe during the period 1960-2100. Therefore will the change in climatic parameters be observed using the climate data created with REMO model. This is done in the start file: "EU_REMO_mapping_start_CC.m". This example will not use a state-space model, but will extract climatic data from the climate file and evaluate mean values.

The following climate data has been analysed for this research: The air temperature [°C], the relative humidity [%], the wind speed [m/s] and the horizontal global solar radiation [W/m²]. The mean value of the climate data for three periods is mapped in an image of Europe. Also is the difference in between the mean values of 'the past' and 'the near future' and the difference between the mean values of 'the past' and 'the far future'.

In figures 9 and 10 has the mean temperature of 'the past' been subtracted from the mean temperatures of 'the near future' and 'the far future'. In figure 9 is shown that a temperature rise of ± 0.0 °C to ± 2.7 °C is expected. In figure 10 is shown that a temperature rise of ± 0.4 °C to ± 4.0 °C is expected. These images also show in which region of Europe the temperature rise will occur the most. For the period 'the near future' is expected that Eastern Europe will have a more increased temperature than Western Europe. The most northern region of Europe and the most southern region of Europe are expected to  have an even higher temperature rise. The same trend is expected for the period 'the far future', but with larger mean air temperature differences between regions.

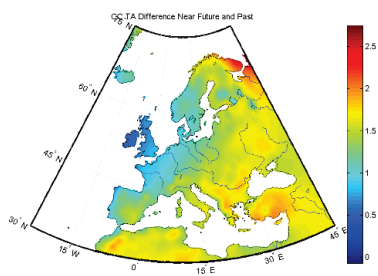

**Figure 9: The difference in mean air temperature between the periods [2020-2050] – [1960-1990].**

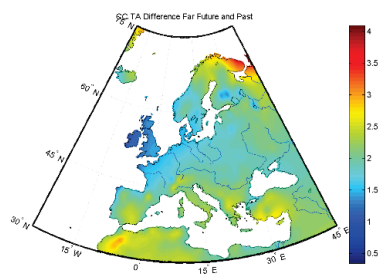

**Figure 10: The difference in mean air temperature between the periods [2090-2100] – [1960-1990].**

In figures 11 and 12 has the mean relative humidity of 'the past' been subtracted from the mean relative humidity of 'the near future' and 'the far future'. In figure 11 is shown that a relative humidity rise of ± -3.5 °C to ± 2.4 °C is expected. In figure 12 is shown that a relative humidity rise of ± -6 °C to ± 3.5 °C is expected. These images also show in which region of Europe the temperature rise will occur the most. For both periods is a decrease in relative humidity expected in the south of Europe. In most parts will the relative humidity not change much.

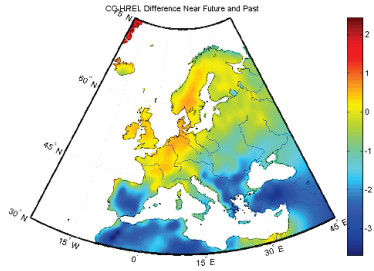

**Figure 11: The difference in mean relative humidity between the periods [2020-2050] − [1960-1990].**

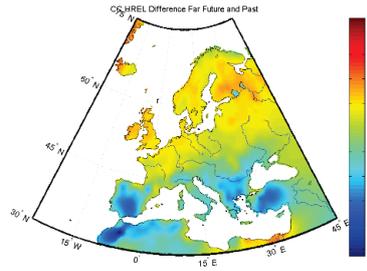

**Figure 12: The difference in mean relative humidity between the periods [2070-2100] − [1960-1990].**

In figures 13 and 14 has the mean wind speed of 'the past' been subtracted from the mean relative humidity of 'the near future' and 'the far future'. Both figures show that the mean wind speed won't change much the next hundred years. For 'the near future' the increase of the wind speed is expected to be ± -0.19 m/s to ± 0.14 m/s and for 'the far future' it is expected to be ± -0.28 m/s to ± 0.17 m/s.

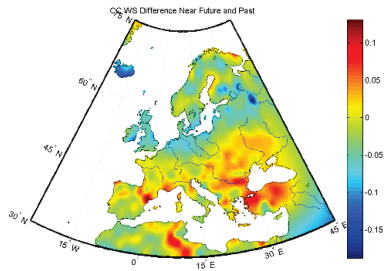

**Figure 13: The difference in mean wind speed between the periods [2020-2050] − [1960-1990].**

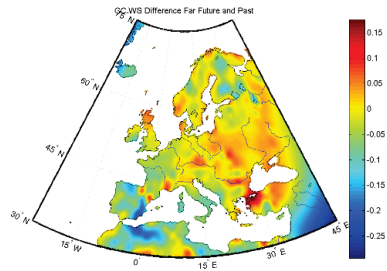

**Figure 14: The difference in mean wind speed between the periods [2070-2100] − [1960-1990].**

The differences in mean horizontal global solar radiation values are only observable when the mean values of 'the past' are subtracted from the mean values of 'the near future' and 'the far future'. This is shown in figure 15 and figure 16. The mean of the horizontal global solar radiation is expected vary slightly from the period 'the past'. The horizontal global solar radiation is expected to increase with ±-12W/m$^2$ tot ±2.5W/m$^2$ in 'the near future'. For 'far future' an increase of ±-12W/m$^2$ tot ±4W/m$^2$ is expected.

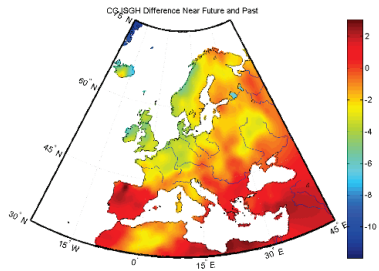 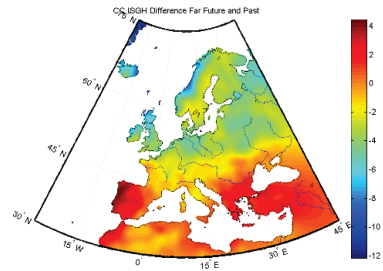

**Figure 15: The difference in mean horizontal global solar radiation between the periods [2020-2050] – [1960-1990].**

**Figure 16: The difference in mean horizontal global solar radiation between the periods [2070-2100] – [1960-1990].**

## 7.2. HVAC system

The performance of a HVAC system as described by C.P. Underwood in HVAC controls systems, has been mapped. This is done in the start file: "EU_REMO_mapping_start_HVAC.m". In figure 17 the HVAC system is shown and in figure 18 an schematic representation has been given. In the system there are three heat flows, one heat exchanger and five states.

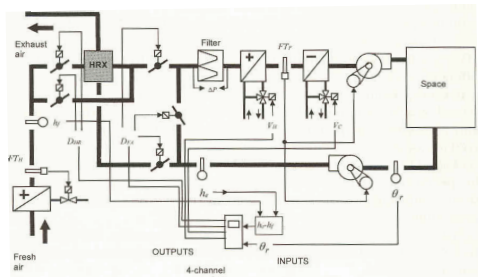

**Figure 17: HVAC system from HVAC controls systems, by C.P. Underwood**

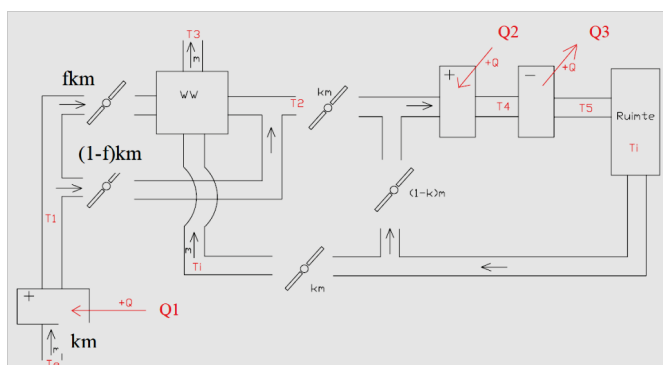

**Figure 18: Schematic representation including air temperatures and heat flows**

The following ODE's can be made to describe the HVAC system. With $T_i$ as the room temperature, $T_e$ as the outdoor temperature, $Q_{1,2,3}$ as heat flows and $Q_{ww}$ as the heat flow of the heat exchanger.

**Equations 2: The ordinary differential equations for the HVAC system.**

$$C_1 \frac{dT_1}{dt} = k \cdot \dot{m} \cdot c \cdot T_e + Q_1 - k \cdot \dot{m} \cdot c \cdot T_1$$

$$C_2 \frac{dT_2}{dt} = k \cdot \dot{m} \cdot c \cdot T_2 - k \cdot \dot{m} \cdot c \cdot T_2 + Q_{ww}$$

$$C_3 \frac{dT_3}{dt} = k \cdot \dot{m} \cdot c \cdot T_i - k \cdot \dot{m} \cdot c \cdot T_3 - Q_{ww}$$

$$C_4 \frac{dT_4}{dt} = k \cdot \dot{m} \cdot c \cdot T_2 + Q_2 + (1-k) \cdot \dot{m} \cdot c \cdot T_i - \dot{m} \cdot c \cdot T_4$$

$$C_5 \frac{dT_5}{dt} = \dot{m} \cdot c \cdot T_4 - Q_3 - \dot{m} \cdot c \cdot T_5$$

With

$$Q_{ww} = K \cdot \left( \frac{T_i + T_3}{2} - \frac{T_1 + T_2}{2} \right)$$

The ODE's can be rewritten to replace $Q_{ww}$. This is necessary due to the fact that $Q_{ww}$ is dependent of the temperatures of state 1, 2 and 3. Only the ODE's for state 2 and 3 need to be rewritten. These rewritten ODE's are shown in

Equations 3.

**Equations 3: The rewritten ordinary differential equations for the HVAC system state 2 and 3.**

$$C_2 \frac{dT_2}{dt} = k \cdot \dot{m} \cdot c \cdot T_2 - k \cdot \dot{m} \cdot c \cdot T_2 + \frac{K}{2} \cdot T_i + \frac{K}{2} \cdot T_3 - \frac{K}{2} \cdot T_1 - \frac{K}{2} \cdot T_2$$

$$C_3 \frac{dT_3}{dt} = k \cdot \dot{m} \cdot c \cdot T_i - k \cdot \dot{m} \cdot c \cdot T_3 - \frac{K}{2} \cdot T_i - \frac{K}{2} \cdot T_3 + \frac{K}{2} \cdot T_1 + \frac{K}{2} \cdot T_2$$

The following state-space matrix model can be created from the ODE's:

$$x(t) = \begin{bmatrix} T_1 \\ T_2 \\ T_3 \\ T_4 \\ T_5 \end{bmatrix}, \quad u(t) = \begin{bmatrix} T_e \\ T_i \\ Q_1 \\ Q_2 \\ Q_3 \end{bmatrix}$$

$$A = \begin{bmatrix} \dfrac{-k \cdot \dot{m} \cdot c}{C_1} & 0 & 0 & 0 & 0 \\[2mm] \dfrac{k \cdot \dot{m} \cdot c - K/2}{C_2} & \dfrac{-k \cdot \dot{m} \cdot c - K/2}{C_2} & \dfrac{K/2}{C_2} & 0 & 0 \\[2mm] \dfrac{K/2}{C_3} & \dfrac{K/2}{C_3} & \dfrac{-k \cdot \dot{m} \cdot c - K/2}{3} & 0 & 0 \\[2mm] 0 & \dfrac{k \cdot \dot{m} \cdot c}{C_4} & 0 & \dfrac{-\dot{m} \cdot c}{C_4} & 0 \\[2mm] 0 & 0 & 0 & \dfrac{\dot{m} \cdot c}{C_5} & \dfrac{-\dot{m} \cdot c}{C_5} \end{bmatrix}$$

$$B = \begin{bmatrix} \dfrac{k \cdot \dot{m} \cdot c}{C_1} & 0 & \dfrac{1}{C_1} & 0 & 0 \\[2mm] 0 & \dfrac{k/2}{C_2} & 0 & 0 & 0 \\[2mm] 0 & \dfrac{k \cdot \dot{m} \cdot c - K/2}{C_3} & 0 & 0 & 0 \\[2mm] 0 & \dfrac{(1-k) \cdot \dot{m} \cdot c}{C_4} & 0 & \dfrac{1}{C_4} & 0 \\[2mm] 0 & 0 & 0 & 0 & \dfrac{-1}{C_5} \end{bmatrix}, C = \begin{bmatrix} 1 & 0 & 0 & 0 & 0 \\ 0 & 1 & 0 & 0 & 0 \\ 0 & 0 & 1 & 0 & 0 \\ 0 & 0 & 0 & 1 & 0 \\ 0 & 0 & 0 & 0 & 1 \end{bmatrix}, D = \begin{bmatrix} 0 & 0 & 0 & 0 & 0 \\ 0 & 0 & 0 & 0 & 0 \\ 0 & 0 & 0 & 0 & 0 \\ 0 & 0 & 0 & 0 & 0 \\ 0 & 0 & 0 & 0 & 0 \end{bmatrix}$$

Matrices C and D give a matrices which result in the output as the temperatures of the five states of the system.

The constants shown in Table 1 are used to calculated the performance of the HVAC system.

**Table 1: Constants used for the HVAC model**

| Parameter | Constant | Value |
|---|---|---|
| Specific heat capacity | c | 1005 J/(kg·K) |
| Density | ρ | 1.2 kg/m³ |
| Volume 1 | $V_1$ | 5 m³ |
| Volume 2 | $V_2$ | 5 m³ |
| Volume 3 | $V_3$ | 5 m³ |
| Volume 4 | $V_4$ | 5 m³ |
| Volume 5 | $V_5$ | 5 m³ |
| Mass flow | ṁ | 0.2 kg/s |
| Valve 1 | k | 1 |
| Valve 2 | f | 1 |
| Constant heat exchanger | K | 200 |
| Heat flow 1 | $Q_1$ | 500 W |
| Heat flow 2 | $Q_2$ | 2000 W |
| Heat flow 3 | $Q_3$ | 500 W |
| Indoor temperature | $T_i$ | 22 °C |
| Heat capacity | $C_n$ | c·ρ·Vn [J/kg] |

The state-space model returns as output a matrix of the temperatures of the five states of the model. These temperatures can be used to evaluate the performance of the system. For this model the heat exchange between the HVAC system and the room will be evaluated. Therefore only the temperature of state 5 of the HVAC system is used. The performance is given in equation 4:

**Equation 4: formula of the performance [W].**

$$P = \dot{m} \cdot c \cdot (T_5 - T_i)$$

This will give a performance for every hour. For the purpose of mapping the collection of hourly performance values needs to be reduced to one single value. Therefore the mean value of all the performance values will be used to evaluate the performance of the system for each location.

In figures 19 and 20 is shown that the mean performance values increase for whole Europe. Central Europe has a smaller increase in performance than the Eastern Europe. In 'the Far Future' is the increase in performance larger.

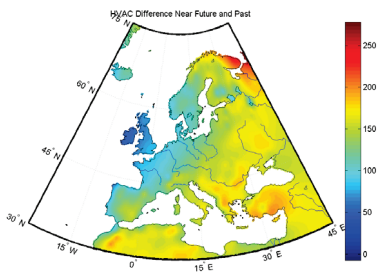

**Figure 19: The difference in performance of the HVAC system between the periods [2020-2050] – [1960-1990].**

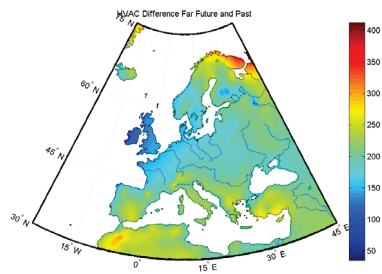

**Figure 20: The difference in performance of the HVAC system between the periods [2070-2100] – [1960-1990].**

### 7.3. Concrete brickwork solar collector

The concrete brickwork solar collector described by F.T. van Schie [7] has been used as an example for this research. The images, the ODE's and the state-space matrices all originate from the work of F.T. van Schie. The performance indicators of F.T. van Schie have been reduced to a single performance indicator to reduce the output, due to each performance indicator creating 5 maps. This is done in the start file: "EU_REMO_mapping_start_SC.m".

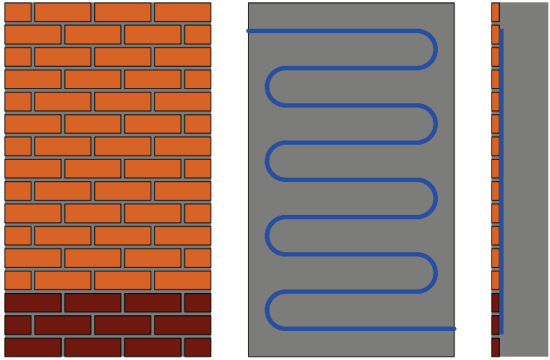

**Figure 21: Visual representation of the concrete brickwork solar collector, by F.T. van Schie [7]**

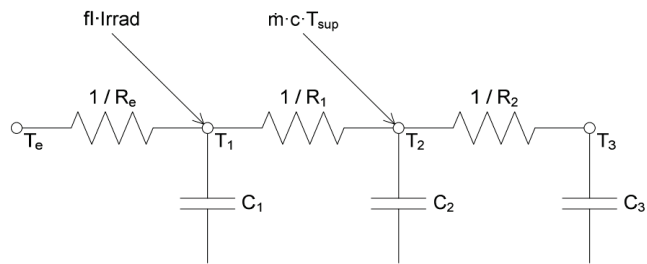

**Figure 22: Thermal network of the concrete brickwork solar collector, by F.T. van Schie [7]**

The system has three states and is influenced by the climate air temperature ($T_e$) and solar radiation (Irrad). The states are also influenced by the temperature of the supply water ($T_{sup}$). This temperature is assumed to be constant.

The thermal network in figure 22 can be translated in the following three ODE's shown in equations 5:

**Equations 5: The ordinary differential equations for the solar collector system.**

$$C_1 \frac{dT_1}{dt} = h \cdot S \cdot (T_e - T_1) - \frac{T_1 - T_2}{R_1} + \propto \cdot S \cdot Irrad$$

$$C_2 \frac{dT_2}{dt} = \dot{m} \cdot c \cdot (T_{sup} - T_2) + \frac{T_1 - T_2}{R_1} - \frac{T_2 - T_3}{R_2}$$

$$C_3 \frac{dT_3}{dt} = \frac{T_2 - T_3}{R_2}$$

The set of ODE's can be translated into the following state-space matrix, with the constants shown in table 2:

$$x(t) = \begin{bmatrix} T_1 \\ T_2 \\ T_2 \end{bmatrix}, \quad u(t) = \begin{bmatrix} T_e \\ T_{sup} \\ Irrad \end{bmatrix}$$

$$A = \begin{bmatrix} \dfrac{-h \cdot S - 1/R_1}{C_1} & \dfrac{1/R_1}{C_1} & 0 \\[4mm] \dfrac{1/R_1}{C_2} & \dfrac{-\dot{m} \cdot c - 1/R_1 - 1/R_2}{C_2} & \dfrac{1/R_2}{C_2} \\[4mm] 0 & \dfrac{1/R_2}{C_3} & \dfrac{-1/R_2}{C_3} \end{bmatrix}$$

$$B = \begin{bmatrix} \dfrac{h \cdot S}{C_1} & 0 & \dfrac{\alpha \cdot S}{C_1} \\[3mm] 0 & \dfrac{\dot{m} \cdot c}{C_2} & 0 \\[3mm] 0 & 0 & 0 \end{bmatrix}, C = \begin{bmatrix} 0 & 0 & 0 \\ 0 & \dot{m} \cdot c & 0 \\ 0 & 0 & 0 \end{bmatrix}, D = \begin{bmatrix} 0 & 0 & 0 \\ 0 & -\dot{m} \cdot c & 0 \\ 0 & 0 & 0 \end{bmatrix}$$

**Table 2: Constants used for the brickwork solar collector model**

| Parameter | Constant | Value |
|---|---|---|
| Surface solar collector | S | 2 m² |
| Specific heat capacity | c | 4200 J/(kg·K) |
| Heat capacity 1 | $C_1$ | 100 000 J/K |
| Heat capacity 2 | $C_2$ | 15 000 J/K |
| Heat capacity 3 | $C_3$ | 300 000 J/K |
| Heat transfer coefficient | h | 25 W/m² |
| Mass flow | $\dot{m}$ | 0.016 kg/s |
| Heat resistance masonry | $R_1$ | 0.1 K/W |
| Heat resistance wall | $R_2$ | 3.0 K/W |
| Absorption coefficient | α | 0.9 |
| Temperature supply water | $T_{sup}$ | 10 °C |

The state-space model results in an output y(t) defined as:

$$y(t) = \dot{m} \cdot c \cdot T_2 - \dot{m} \cdot c \cdot T_{supp}$$

The added heat from the solar collector to the heating medium is modelled in this state-space model. The matrices C and D are constructed in a way to obtain this specific output. This is more direct than the example of the HVAC system. The HVAC system had an output that resulted in the temperatures of the different states. Additional calculations were necessary to result in a performance indicator. For the brickwork solar collector the performance indicator is directly the output. The only additional calculation needed is calculating one mean value from the many hourly values.

The performance indicator is computed for each time moment. The mean of the results per location has been mapped in an image of Europe.

In figures 23 and 24 are the mean performance values of 'the past' subtracted from the other two time periods. This shows that for whole Europe the performance will increase of the solar collector. The regions near the North Sea will have a minimal increase in performance. Eastern Europe and North Africa show a higher increase in performance. The 'Far Future' shows the same trend as the 'Near Future', but the increase in performance is larger.

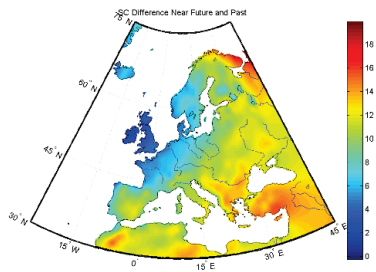

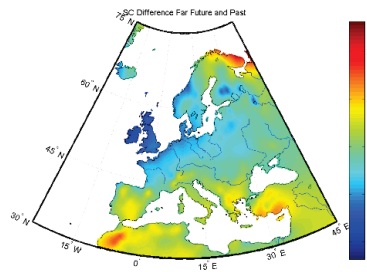

**Figure 23: The difference in performance of the solar collector between the periods [2020-2050] – [1960-1990].**

**Figure 24: The difference in performance of the solar collector between the periods [2070-2100] – [1960-1990].**

# 8. Sources

[1]   Li DHW, Yang L & Lam JC (2012). *Impact of climate change on energy use in the built environment in different climate zones - A review.* Energy 42 (2012) 103-112

[2]   Wilde P de, Coley D (2011). *The implications of a changing climate for buildings.* Building and Environment 55 (2012) 1–7

[3]   Xu P, Huang YJ, Miller N, Schlegel N & Shen P (2012). *Impacts of climate change on building heating and cooling energy patterns in California.* College of Mechanical Engineering, Tongji University, Shanghai, China. Energy 44 (2012) 792-804

[4]   Wachsmuth J, Blohm A, Göβling-Reisemann S, Eickemeier T, Ruth M, Gasper R & Stührmann S (2013).  *How will renewable power generation be affected by climate change? The case of a Metropolitan Region in Northwest Germany.* Energy 58 (2013) 192-201

[5]   Gaetani M, Huld T, Vignati E, Monfortio-Ferrario F, Dosio A & Raes F (2014). *The near future availability of photovoltaic energy in Europe and Africa in climate-aerosol modelling experiments.* RenewableandSustainableEnergyReviews38(2014)706–716

[6]   Climate for Culture. Climate for Culture. Retrieved from: http://www.climateforculture.eu/index.php?inhalt=project.overview (2014).

[7]   Schie FT van (2013). *A Combination Of State Space and Mapping Functions In MATLAB: Evaluation tool for climate related cases in the Netherlands.* Eindhoven, Technische Universiteit Eindhoven.

[8]     Schijndel AWM van & Schellen HL (2013). *The simulation and mapping of building performance indicators based on european weather stations.* Frontiers of Architectural Research, 2(2), 121-133.

[9]     Climate for Culture. Climate Change Modelling. Retrieved from: http://www.climateforculture.eu/index.php?inhalt=project.climatechange (2014).